# Temperature-independent thermal radiation

Alireza Shahsafi[1*], Patrick Roney[1*], You Zhou[2], Zhen Zhang[3], Yuzhe Xiao[1], Chenghao Wan[1,4], Raymond Wambold[1], Jad Salman[1], Zhaoning Yu[1,5], Jiarui Li[6], Jerzy T. Sadowski[7], Riccardo Comin[6], Shriram Ramanathan[3], and Mikhail A. Kats[1,4,5†]

[1]Department of Electrical and Computer Engineering, University of Wisconsin – Madison, Madison, Wisconsin
[2]School of Engineering and Applied Sciences, Harvard University, Cambridge, Massachusetts
[3]School of Materials Engineering, Purdue University, West Lafayette, Indiana
[4]Materials Science and Engineering, University of Wisconsin – Madison, Madison, Wisconsin
[5]Department of Physics, University of Wisconsin – Madison, Madison, Wisconsin
[6]Department of Physics, Massachusetts Institute of Technology, Cambridge, MA
[7]Center for Functional Nanomaterials, Brookhaven National Laboratory, Upton, New York

\* These authors contributed equally
† Correspondence to mkats@wisc.edu

**Abstract:**

Thermal emission is the process by which all objects at non-zero temperatures emit light, and is well-described by the classic Planck, Kirchhoff, and Stefan-Boltzmann laws. For most solids, the thermally emitted power increases monotonically with temperature in a one-to-one relationship that enables applications such as infrared imaging and non-contact thermometry. Here, we demonstrate ultrathin thermal emitters that violate this one-to-one relationship via the use of samarium nickel oxide (SmNiO₃), a strongly correlated quantum material that undergoes a fully reversible, temperature-driven solid-state phase transition. The smooth and hysteresis-free nature of this unique insulator-to-metal (IMT) phase transition allows us to engineer the temperature dependence of emissivity to precisely cancel out the intrinsic blackbody profile described by the Stefan-Boltzmann law, for both heating and cooling. Our design results in temperature-independent thermally emitted power within the long-wave atmospheric transparency window (wavelengths of 8 – 14 µm), across a broad temperature range of ~30 °C, centered around ~120 °C. The ability to decouple temperature and thermal emission opens a new gateway for controlling the visibility of objects to infrared cameras and, more broadly, new opportunities for quantum materials in controlling heat transfer.

**Main text:**

The total amount of power thermally emitted by a surface in free space can be obtained by integrating its spectral radiance—given by Planck's law and an emissivity—over all wavelengths and hemispherical angles [1,2]. Assuming negligible angular dependence of the emissivity and wrapping the angular integral into the blackbody distribution, $I_{BB}(\lambda, T)$, this relationship can be expressed as:

$$A \int_\lambda d\lambda\, \varepsilon(\lambda, T) I_{BB}(\lambda, T) = A \varepsilon_{tot}(T) \sigma T^4, \qquad \text{Eqn. (1)}$$

where $A$ is the surface area, $\varepsilon(\lambda, T)$ is the spectral emissivity, $\lambda$ is the free-space wavelength, $T$ is the temperature, and $\sigma$ is the Stefan-Boltzmann constant. The total emissivity, $\varepsilon_{tot}(T)$, can have a gradual temperature dependence even if the spectral emissivity has no such dependence, due to the integration of $\varepsilon(\lambda) I_{BB}(\lambda, T)$ [3]; nevertheless, this dependence is usually dwarfed by the $T^4$ term, and so $\varepsilon_{tot}$ can often be considered to be approximately constant. Thus, the Stefan-Boltzmann law yields a one-to-one mapping between the temperature of an object and the emitted power, resulting in the conventional wisdom that hotter objects emit more light [Fig. 1(a, c)], and enabling applications such as infrared imaging and non-contact thermometry [4,5].

This assumption of a near-constant emissivity must be re-examined for thermal emitters comprising materials whose optical properties can be widely tunable with temperature (i.e., thermochromics). For example, an emissivity that increases with temperature can result in emitted power growing faster than $T^4$ [6,7], and an emissivity that rapidly decreases with temperature can overwhelm and reverse the slope of the typical Stefan-Boltzmann curve [8,9].

Here, we show that it is possible to achieve a complete breakdown of the conventional one-to-one mapping between the temperature and the thermally emitted power, $P$. A thermal-emission coating with this unique property can serve as a radiator that outputs a fixed amount of heat irrespective of its temperature, and can conceal differences in temperature across an object from infrared imagers. This condition can be written as $\partial P/\partial T = 0$, and occurs when $\varepsilon_{tot} = \gamma T^{-4}$, where $\gamma$ is a constant with units of K$^4$. A surface with $\varepsilon_{tot}$ that fits this form over some temperature range is henceforth referred to as a *zero-differential thermal emitter* (ZDTE). Achieving ZDTE behavior using real materials is extremely challenging: the necessary rate of change of the emissivity with temperature is much larger than what can be attained using conventional materials [e.g., with band semiconductors such as silicon, via temperature-dependent population of electrons to the conduction band, Fig. 1(b)], but is smaller than that of materials with abrupt phase transitions (e.g., vanadium dioxide [8,10]). Furthermore, this condition is only possible for a *hysteresis-free* temperature dependence of the emissivity, otherwise the ZDTE condition may only be satisfied during either heating or cooling, but not both.

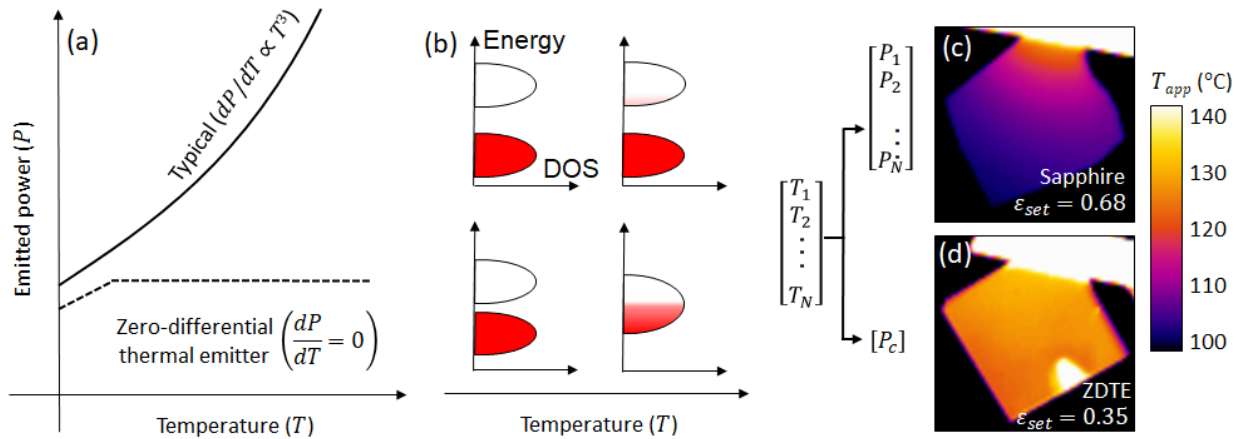

**Figure 1**. Comparison between a typical thermal emitter and a zero-differential thermal emitter (ZDTE). **(a, b)** For a typical emitter, for example comprising a semiconductor or insulator (cartoon band diagram in (b), top), any change in emission from a temperature-dependent change in materials properties is dwarfed by the $T^4$ dependence in the Stefan-Boltzmann law. Conversely, a ZDTE decouples temperature and thermal radiation over some temperature range, and thus can only be made from a material with a very strong temperature dependence. In our implementation, we use the hysteresis-free insulator-to-metal phase transition in samarium nickelate (SmNiO$_3$) to achieve this behavior (b, bottom). **(c, d)** Long-wave infrared images of samples mounted to hang off the edge of a heater stage, such that a temperature gradient is established from hot (top) to cold (bottom). **(c)** a reference sample with a constant emissivity—in this case, a sapphire wafer—and **(d)** a ZDTE based on SmNiO$_3$. The color bar encodes the *apparent* temperature,

obtained by assuming a particular set emissivity, $\varepsilon_{set}$, which was chosen such that the sample region just below the heat stage appeared to be at 130 °C, which is the actual temperature at that point (see more discussion in *Methods*). For sapphire, there is a one-to-one relationship between temperature and thermally emitted power. Conversely, the ZDTE exhibits a constant emitted power over a range of temperatures, here approximately 100 – 135 °C.

Here, we demonstrate ZDTE in the 8 – 14 µm atmospheric-transparency window [11] using samarium nickel oxide (SmNiO$_3$), a correlated perovskite that features strong yet relatively gradual evolution of its optical properties over the temperature range of ~40 to ~140 °C, resulting from a fully reversible and hysteresis-free thermally driven IMT [12] [13] [14] [15] [16]. The thermal IMT in SmNiO$_3$ is due to charge disproportionation in the Ni site and involves subtle changes in the Ni-O-Ni bond angle [12,16]. In our SmNiO$_3$ films (see *Methods*), this thermally driven transition is reversible over many cycles, and has essentially no hysteresis in both electrical [Fig. 2(a)] and optical measurements [Fig. 2(b)], in stark contrast to many other materials with strong IMTs, e.g., vanadium dioxide [17]. Hysteresis-free IMTs can be found in rare-earth nickelates with high phase transition temperatures, where negligible or complete absence of hysteresis may be due to the decoupling of the IMT with antiferromagnetic ordering, and faster phase-transformation kinetics at higher temperatures [18] [19]. The unique nature of this IMT is also directly observed in our spatially-resolved X-ray absorption spectroscopy (XAS) maps across the thermal transition [inset of Fig. 2(a)], which demonstrate smooth variation with temperature (i.e., the absence of any metallic/insulating domain texture at any temperature) down to a ~20 nm length scale, including for temperatures deep within the IMT; see *Methods* and *Supplementary* 4 for details. No spatial features other than detector noise were observed. The trend in these spatial maps (including additional data in *Supplementary* 4) suggests a smooth crossover from the insulating limit to the metallic one, which is accompanied by a homogeneous phase landscape. The *Supplementary Information* includes experimental data showing the stability over many cycles (*Supplementary* 1) and x-ray diffraction measurements that further help explain the transition behavior (*Supplementary* 2).

To enable design of thermal emitters using SmNiO$_3$, we performed temperature-dependent variable-angle spectroscopic ellipsometry over the 2 – 16 µm wavelength range, through the entire range of the phase transition [Fig. 2 (c, d)]. The resulting complex refractive-index data is consistent with the film becoming gradually more metallic from room temperature to ~140 °C. We note that while gradual transitions are generally considered to be less useful than abrupt transitions for electronic and optical switching technologies, here the gradual and hysteresis-free nature of the IMT in SmNiO$_3$ is essential for the realization of ZDTEs.

To minimize fabrication complexity and cost and thus realize robust and large-area ZDTEs, we explored designs based on un-patterned thin films of SmNiO$_3$. We used the temperature-dependent optical properties in Fig. 2(c, d) and well-established optical thin-film calculations [20] to find the necessary combination of thickness of a SmNiO$_3$ film and a substrate that supports SmNiO$_3$ synthesis (Fig. 3(a) and *Supplementary* 3) to achieve ZDTE over the 8 – 14 µm atmospheric transparency window. In Fig. 3(a), we plotted the calculated temperature derivative of the emitted integrated radiance for several thicknesses of SmNiO$_3$ on sapphire, which indicates that ZDTE can be achieved for SmNiO$_3$ thickness of ~150 nm or greater. The result does not change much for SmNiO$_3$ films thicker than ~250 nm, indicating that the optical properties of the substrate do not affect the emissivity, making SmNiO$_3$ a versatile surface coating that can be utilized for scalable technologies.

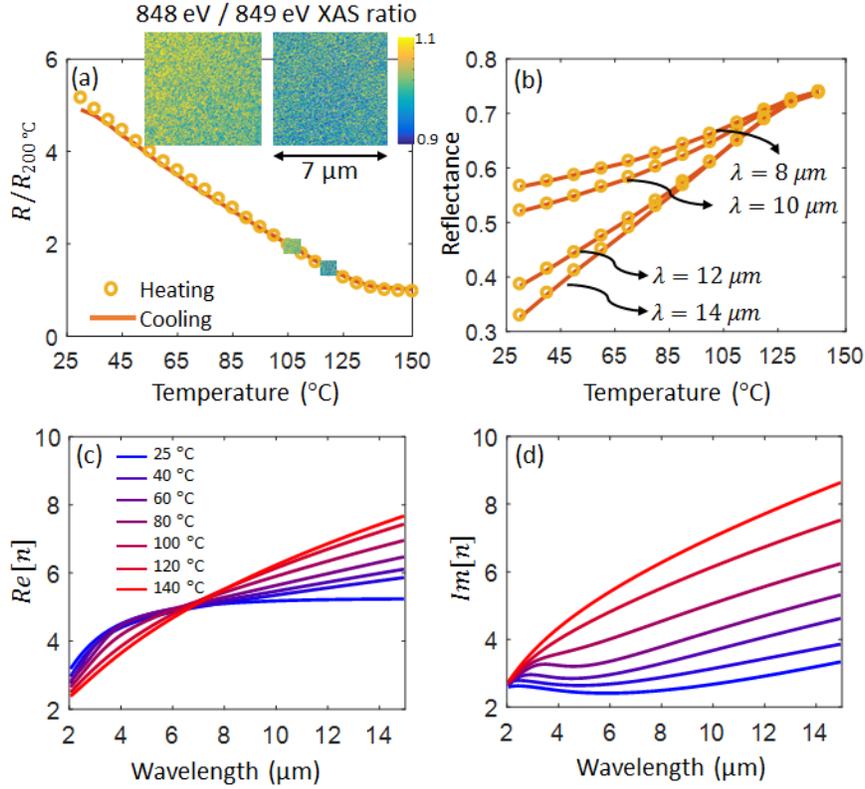

**Figure 2**: Hysteresis-free insulator-to-metal phase transition in SmNiO$_3$. **(a)** Normalized temperature-dependent electrical resistance of our SmNiO$_3$ thin film grown on a sapphire substrate and **(b)** mid-infrared reflectance at several representative wavelengths, during both heating and cooling, showing the hysteresis-free nature of the IMT in SmNiO$_3$. The insets in (a) are nanoscale x-ray absorption (XAS) maps at 105 and 120 °C, where the ratio of x-ray absorption at 848 eV to that at 849 eV is plotted as an indication of the metallic/insulating properties; no features other than detector noise are observed, indicating a gradual transition with no observable domain texture. **(c, d)** Temperature-dependent (c) real and (d) imaginary parts of the complex refractive index of the SmNiO$_3$ film, as a function of wavelength across the mid infrared, extracted using spectroscopic ellipsometry.

Our fabricated planar device consists of a ~220-nm SmNiO$_3$ film grown on a *c*-plane sapphire substrate (Fig. 3(a) inset; see *Methods* for details), from which we measured the emissivity and the resulting thermally emitted spectral radiance. Because the resulting structure is opaque in our wavelength region of interest (due to the optical losses in SmNiO$_3$ as well as in sapphire [21]) and flat on the scale of the wavelength, Kirchhoff's law can be used to calculate the normal-direction emissivity $\varepsilon_N(\lambda, T)$ from normal-incidence reflection measurements: $\varepsilon_N(\lambda, T) = 1 - R_N(\lambda, T)$ (see *Methods* for details). We confirmed this result by measuring the thermal emission directly, normalizing to a laboratory blackbody consisting of a vertically-oriented 0.1-mm-tall carbon nanotube forest (see *Methods*). Care was taken to isolate the sample thermal emission from the thermal background radiated by the various components of our instrument [22] (see *Supplementary* 5). The results obtained from these two measurements are in excellent agreement [Fig. 3(b)].

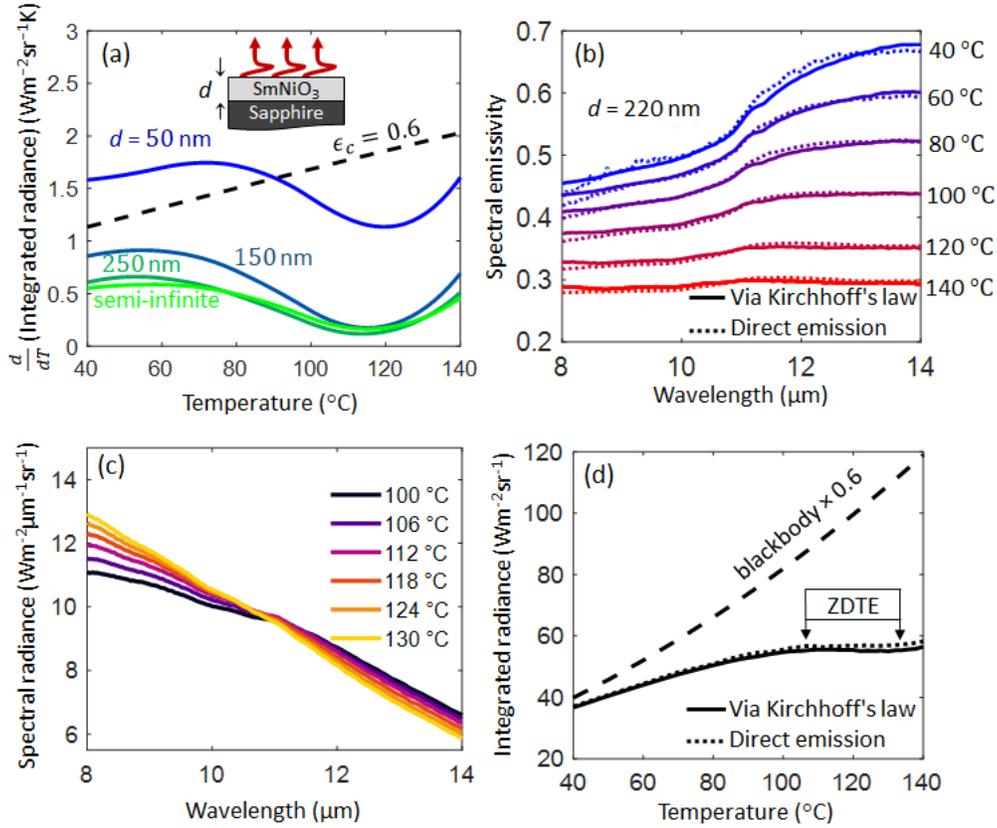

**Figure 3**: Zero-differential thermal emission. **(a)** Calculated temperature derivative of the emitted radiance, integrated over the 8-14 µm atmospheric transparency window, of an SmNiO$_3$ film with thicknesses $d$ from 50 nm to infinity, on a semi-infinite sapphire substrate. **(b)** Measured wavelength- and temperature-dependent emissivity of our ZDTE, comprising a ~220-nm film of SmNiO$_3$ on a sapphire substrate, via direct emission (dotted) and Kirchhoff's law using reflection measurements (solid). **(c)** The temperature-dependent spectral radiance of the ZDTE, which is the product of the spectral emissivity in (b) and the Planck distribution. **(d)** Thermally emitted radiance of our ZDTE, integrated over 8-14 µm, compared to that of a black body.

We integrated the measured spectral radiance [Fig. 3(c)] over the 8–14 µm window to obtain the total thermally emitted integrated radiance as a function of temperature, which showed the desired ZDTE effect within a ~30 °C temperature window centered around ~120 °C [Fig. 3(d)]. Away from the center of the phase transition of SmNiO$_3$, *i.e.*, below ~80 and above ~140 °C, the radiance vs. temperature profile becomes monotonic, as expected for a typical non-thermochromic thermal emitter. We note that the ZDTE effect is quite robust for different angles of emission (*Supplementary* 6).

The presence of the zero-differential region has profound implications for infrared imaging and control of infrared visibility. To demonstrate this, we performed a model experiment where two samples—our ZDTE, and a reference sapphire wafer—were mounted on a temperature-controlled chuck such that only a corner was touching the chuck and most of the sample was suspended in air, resulting in a temperature gradient from ~140 directly on top of the chuck to ~105 °C at the corner of the suspended area. When imaged with a long-wave infrared (LWIR) camera, the gradient is readily observable on the sapphire reference [Fig. 1(b)], but almost completely disappears on our SmNiO$_3$-based ZDTE [Fig. 1(d)]. Note that the bright spot in the bottom side of the ZDTE sample corresponds to a region with no $SmNiO_3$ (i.e., it is simply sapphire). This is the area covered by the clip used to mount the sample in the sputtering chamber. The apparent temperature difference across the samples based on the camera image was ~34 °C for the sapphire, and ~9 °C for the ZDTE. The same phenomenon is observed in Fig. 4, where we show the temperature evolution of the infrared appearance of $SmNiO_3$-based ZDTEs compared to our laboratory blackbody reference (carbon nanotube forest), and sapphire and fused-silica wafers. Note that we numerically analyzed the temperature drop across the thickness of the sapphire wafer (i.e., from the heat stage to the surface of the sample) and found that it is less than 0.1 °C (*Supplementary 7*).

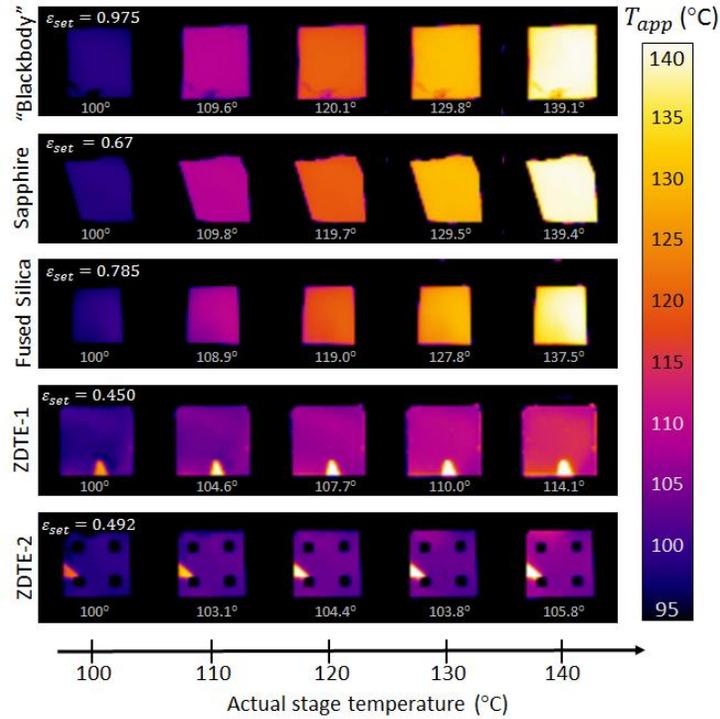

**Figure 4**: Long-wave infrared images of samples held at temperatures from 100 to 140 °C. The emissivities of the laboratory blackbody (carbon nanotube forest), sapphire wafer, and fused $SiO_2$ wafer do not change appreciably over this temperature range. The emissivities of our $SmNiO_3$-based ZDTEs change as a function of temperature, and thus effectively mask the temperature differences from the camera. The apparent temperature is plotted [like in Fig. 1(c, d)], with $\varepsilon_{set}$ for each sample selected such that for a stage temperature of 100 °C, the infrared camera returned this value as the temperature reading. The dark squares on the bottom row are metal electrodes that were used for the resistance measurements in Fig. 2(a).

We note that the presence of zero-differential emission does not necessarily guarantee completely temperature-independent infrared signatures. For example, slight differences can still be observed in the emission from our ZDTEs at 110 °C vs. 140 °C in Fig. 4, resulting from a combination of effects that include imperfect cancelation of the blackbody curve and the change of the reflected light from the environment, since a change in emissivity coincides with a change in reflectance. The latter can be compensated during the design of ZDTEs by considering both the emitted and reflected light, assuming a particular background temperature.

In conclusion, we demonstrated that the typical one-to-one relationship between temperature and thermal radiation can be severed using zero-differential thermal emitter (ZDTE) coatings based on SmNiO$_3$, a quantum material featuring a hysteresis-free thermally driven insulator-to-metal transition. The temperature range of the zero-differential emission effect can be widely tuned by strain, doping, or tilt control of SmNiO$_3$, which can shift the transition range to room temperature and even below [15][23][24][25]. For example, alloys of samarium and neodymium nickelates can have IMT temperatures roughly between -100 °C and 100 °C depending on the specific composition [26], which may enable ZDTE design over a wide range of temperatures (*see Supplementary 8*). The ability to decouple temperature and thermal radiation with our simple design enables new approaches to conceal heat signatures over large areas, for example for wearable personal privacy technologies, and also has implications for thermal management in space. More broadly, this demonstration can motivate new areas of inquiry for quantum materials that possess highly tunable electronic structures.

**Methods**

Materials synthesis: The SmNiO$_3$ films were grown using magnetron sputtering. The sputtering power was set to 90 W DC for the Ni target and 170 W RF for the Sm target. The total pressure during deposition was kept at 5 mTorr under a constant gas flow of 40 sccm for Ar and 10 sccm for O$_2$. The as-deposited films were subsequently transferred into a home-built high-pressure chamber and annealed in 1400-psi oxygen gas at 500 °C for 24 hours to form the perovskite phase. The electrical resistance of SmNiO$_3$ films [Fig. 2(a)] was measured on a temperature-controlled probe station by sweeping the voltage from −0.1 V to 0.1 V with a Keithley 2635A source meter, with Pt electrodes on top of the SmNiO$_3$ films. The electrodes can be seen in the bottom row of Fig. 4.

Nanoscale X-ray absorption spectroscopy: The x-ray absorption spectroscopy (XAS) / x-ray photoemission electron microscopy (XPEEM) experiments were performed at the XPEEM end station of the Electron Spectro-Microscopy beamline (21-ID-2) at the National Synchrotron Light Source II. The sample was illuminated with a focused monochromatic soft x-ray beam with a photon energy tuned around the Ni-L$_3$ resonance (840 – 850 eV). The temperature was controlled to within 0.1 °C of the target setpoint. All images are drift corrected and the same field of view was recovered at each temperature. All measurements were taken with linear-horizontal polarized light at an incident angle of 73° to the surface normal. The SmNiO$_3$ film used for XAS/XPEEM measurements was deposited epitaxially on an LaAlO$_3$ (001) substrate. The XAS/XPEEM is a spatially resolved probe to measure the Ni valence electronic states during the IMT. The metallic/insulating phases can be distinguished by different XAS spectral shapes. The maps in the inset of Fig. 2(a) are ratios of the absorption coefficients at 848 and 849 eV, which can be used as a proxy for spatial identification of the metallic/insulating phases. This ratio is 1.02 in the insulating phase of SmNiO$_3$ and 0.98 in the metallic phase (see *Supplementary* 4).

Optics measurements: The temperature-dependent complex refractive index of SmNiO$_3$ was measured using a Woollam IR VASE MARK II spectroscopic ellipsometer with a temperature-controlled stage, assuming the film thickness obtained from SEM imaging of the cross section. The fitting was performed using WVASE software. The SmNiO$_3$ film was assumed to be isotropic, whereas anisotropy in the sapphire was included in the fitting model. The reflection measurements were obtained with a Bruker Vertex 70 FTIR and a Hyperion 2000 microscope with a reflective objective (numerical aperture (NA) = 0.4), and a Linkam THMS600 temperature-controlled stage. The direct-emission measurements were obtained with the temperature stage and sample in the FTIR sample compartment, using a parabolic mirror for collection (NA = 0.05). In both direct-emission and Kirchhoff's-law measurements, we used a liquid-nitrogen-cooled mercury-cadmium-telluride (MCT) detector and a potassium bromide (KBr) beam splitter. A gold mirror was used as the reflection reference, whereas a vertically aligned carbon nanotube (CNT) forest on a silicon substrate was used as the emission reference (see *Supplementary* 5 for details about reference calibration and the accounting for background thermal radiation). LWIR imaging was carried out using a FLIR A325sc camera, sensitive to the 7.5 – 13 $\mu$m range.

The color bars of the infrared images in Fig. 1(c, d) and Fig. 4 report the apparent temperature, given some emissivity setting, $\varepsilon_{set}$, in the FLIR camera software. The value of $\varepsilon_{set}$ used for each image is provided directly on the image. In Fig. 1(c, d) and Fig. 4, we selected $\varepsilon_{set}$ such that the apparent temperature corresponded to the actual temperature at some particular point. In Fig. 1(c, d), this point is at the very top of the sample where it just touches the temperature stage; in Fig. 4, this is done for each sample at 100 °C.

Optics calculations: For the calculation of emissivity, we used the transfer-matrix method together with the optical properties from ellipsometry to obtain the absorptivity, which we converted to emissivity using Kirchhoff's law. The emitted spectrum was then calculated by multiplying the spectral emissivity by the Planck distribution at the appropriate temperature, which could then be integrated over the mid-infrared transparent window of 8 to 14 $\mu m$ to obtain the emitted power.

**Data Availability:** The datasets generated during and/or analyzed during the current study are available from the corresponding author on reasonable request.

**Acknowledgements**

MK acknowledges financial support from the Office of Scientific Research (N00014-16-1-2556) and the National Science Foundation (ECCS-1750341). SR acknowledges financial support from the AFOSR (FA9550-16-1-0159). PR was supported by a Critical Skills Master's Fellowship from Sandia National Labs. This research used resources of the Center for Functional Nanomaterials and National Synchrotron Light Source II, which are U.S. DOE Office of Science Facilities, at Brookhaven National Laboratory under Contract No. DE-SC0012704.

**Author Contributions**

MAK conceived of the research. PR and AS carried out the majority of the experiments and data analysis, and are co-equal lead contributors. YZ and ZZ synthesized the SmNiO$_3$ samples, and carried out materials analysis. YX, CW, RW, JS, and ZY contributed to the optical experiments and data analysis, especially with instrument characterization (YX, CW, JS), and custom equipment design (RW). JL, JTS, and RC carried out the XAS/XPEEM measurements. PR, AS, SR, and MAK wrote the manuscript, with input and editing from all co-authors. MAK and SR supervised the project.

**References**

1. Boyd, R. Radiometry and the detection of optical radiation. (1983).
2. Mink, J. *Handbook of Vibrational Spectroscopy*. (Wiley, 2006).
3. Saunders, P. On the effects of temperature dependence of spectral emissivity in industrial radiation thermometry. *High Temp. -- High Press.* **33,** 599–610 (2001).
4. Kaplan, H. *Practical Applications of Infrared Thermal Sensing and Imaging Equipment*. (SPIE, 2007).

5. Vollmer, M. & Möllmann, K.-P. *Infrared Thermal Imaging*. (Wiley-VCH Verlag GmbH & Co. KGaA, 2017). doi:10.1002/9783527693306
6. Benkahoul, M. *et al.* Thermochromic VO2 film deposited on Al with tunable thermal emissivity for space applications. *Sol. Energy Mater. Sol. Cells* **95,** 3504–3508 (2011).
7. Wu, S.-H. *et al.* Thermal homeostasis using microstructured phase-change materials. *Optica* **4,** 1390 (2017).
8. Kats, M. A. *et al.* Vanadium Dioxide as a Natural Disordered Metamaterial: Perfect Thermal Emission and Large Broadband Negative Differential Thermal Emittance. *Phys. Rev. X* **3,** 041004 (2013).
9. Bierman, D. M. *et al.* Radiative Thermal Runaway Due to Negative-Differential Thermal Emission Across a Solid-Solid Phase Transition. *Phys. Rev. Appl.* **10,** 021001 (2018).
10. Joulain, K., Ezzahri, Y., Drevillon, J. & Ben-Abdallah, P. Modulation and amplification of radiative far field heat transfer: Towards a simple radiative thermal transistor. *Appl. Phys. Lett.* **106,** 133505 (2015).
11. Salisbury, J. W. & D'Aria, D. M. Emissivity of terrestrial materials in the 8–14 μm atmospheric window. *Remote Sens. Environ.* **42,** 83–106 (1992).
12. Pérez-Cacho, J., Blasco, J., García, J., Castro, M. & Stankiewicz, J. Study of the phase transitions in SmNiO3. *J. Phys. Condens. Matter* **11,** 405–415 (1999).
13. Jaramillo, R., Ha, S. D., Silevitch, D. M. & Ramanathan, S. Origins of bad-metal conductivity and the insulator–metal transition in the rare-earth nickelates. *Nat. Phys.* **10,** 304–307 (2014).
14. Catalano, S. *et al.* Rare-earth nickelates $R$ NiO $_3$ : thin films and heterostructures. *Reports Prog. Phys.* **81,** 046501 (2018).
15. Ha, S. D., Otaki, M., Jaramillo, R., Podpirka, A. & Ramanathan, S. Stable metal–insulator transition in epitaxial SmNiO3 thin films. *J. Solid State Chem.* **190,** 233–237 (2012).
16. Catalano, S. *et al.* Electronic transitions in strained SmNiO $_3$ thin films. *APL Mater.* **2,** 116110 (2014).
17. Klimov, V. A. *et al.* Hysteresis loop construction for the metal-semiconductor phase transition in vanadium dioxide films. *Tech. Phys.* **47,** 1134–1139 (2002).
18. Catalan, G. Progress in perovskite nickelate research. *Phase Transitions* **81,** 729–749 (2008).
19. Ruppen, J. *et al.* Impact of antiferromagnetism on the optical properties of rare-earth nickelates. *Phys. Rev. B* **96,** 045120 (2017).
20. Kats, M. A. & Capasso, F. Optical absorbers based on strong interference in ultra-thin films. *Laser Photon. Rev.* **10,** 735–749 (2016).
21. Schubert, M., Tiwald, T. E. & Herzinger, C. M. Infrared dielectric anisotropy and phonon modes of sapphire. *Phys. Rev. B* **61,** 8187–8201 (2000).
22. Xiao, Y. *et al.* Measuring Thermal Emission Near Room Temperature Using Fourier-Transform Infrared Spectroscopy. *Phys. Rev. Appl.* **10,** 1 (2019).
23. Xiang, P.-H. *et al.* Room temperature Mott metal-insulator transition and its systematic control in Sm1−xCaxNiO3 thin films. *Appl. Phys. Lett.* **97,** 032114 (2010).
24. Middey, S. *et al.* Physics of Ultrathin Films and Heterostructures of Rare-Earth Nickelates. *Annu. Rev. Mater. Res.* **46,** 305–334 (2016).
25. Liao, Z. *et al.* Metal-insulator-transition engineering by modulation tilt-control in perovskite nickelates for room temperature optical switching. *Proc. Natl. Acad. Sci. U. S. A.* 201807457 (2018). doi:10.1073/pnas.1807457115
26. Girardot, C., Kreisel, J., Pignard, S., Caillault, N. & Weiss, F. Raman scattering investigation across the magnetic and metal-insulator transition in rare earth nickelate R NiO 3 ( R = Sm , Nd ) thin films. *Phys. Rev. B* **3,** 1–7 (2008).

Supplementary information:

# Temperature-independent thermal radiation

Alireza Shahsafi[1*], Patrick Roney[1*], You Zhou[2], Zhen Zhang[3], Yuzhe Xiao[1], Chenghao Wan[1,4], Raymond Wambold[1], Jad Salman[1], Zhaoning Yu[1,5], Jiarui Li[6], Jerzy T. Sadowski[7], Riccardo Comin[6], Shriram Ramanathan[3], and Mikhail A. Kats[1,4,5†]

[1]Department of Electrical and Computer Engineering, University of Wisconsin – Madison, Madison, Wisconsin
[2]School of Engineering and Applied Sciences, Harvard University, Cambridge, Massachusetts
[3]School of Materials Engineering, Purdue University, West Lafayette, Indiana
[4]Materials Science and Engineering, University of Wisconsin – Madison, Madison, Wisconsin
[5]Department of Physics, University of Wisconsin – Madison, Madison, Wisconsin
[6]Department of Physics, Massachusetts Institute of Technology, Cambridge, MA
[7]Center for Functional Nanomaterials, Brookhaven National Laboratory, Upton, New York

* These authors contributed equally
† Correspondence to mkats@wisc.edu

## 1. Temperature-dependent reflectance characterization and cycling tests

To ensure that the optical response of $SmNiO_3$ has no hysteresis, we performed temperature-dependent reflection measurements for a full cycle of heating and cooling [Fig. S1(a-b)] on the same $SmNiO_3$/sapphire sample used for the main text. Various temperature-dependent optical measurements have been performed several times over a period of approximately two years, with no apparent degradation of the samples.

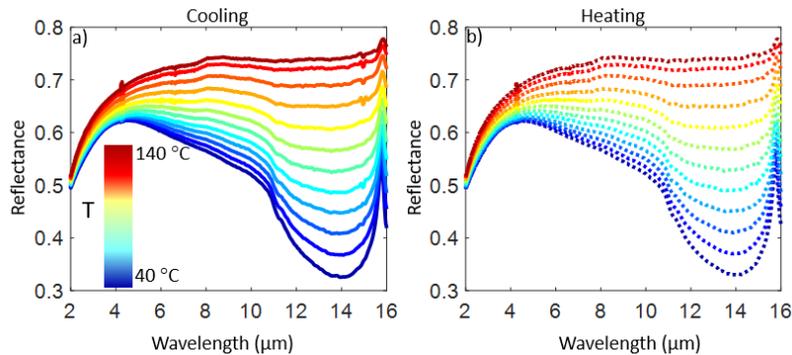

**Figure S1.** Reflection measurements on the sample used for the main text taken under both **(a)** cooling and **(b)** heating. We observe no hysteresis (see also Fig. 2(a, b) in the main text).

To further investigate the stability of our $SmNiO_3$ films, we also analyzed a different sample grown under the same conditions, consisting of a ~200-nm film of $SmNiO_3$ on an $LaAlO_3$ substrate [inset of Fig. S2(a)]. No substantial differences in $SmNiO_3$ quality or degradation was observed between the two substrates ($LaAlO_3$ vs. sapphire). We performed temperature-dependent reflection measurements on this sample and thermally drove it between the two phases every few minutes. The reflectances at different wavelengths [Fig. S2(c-f)] shows that the optical response of the film is stable and has no memory over many cycles, as was expected from the literature [S1-S2].

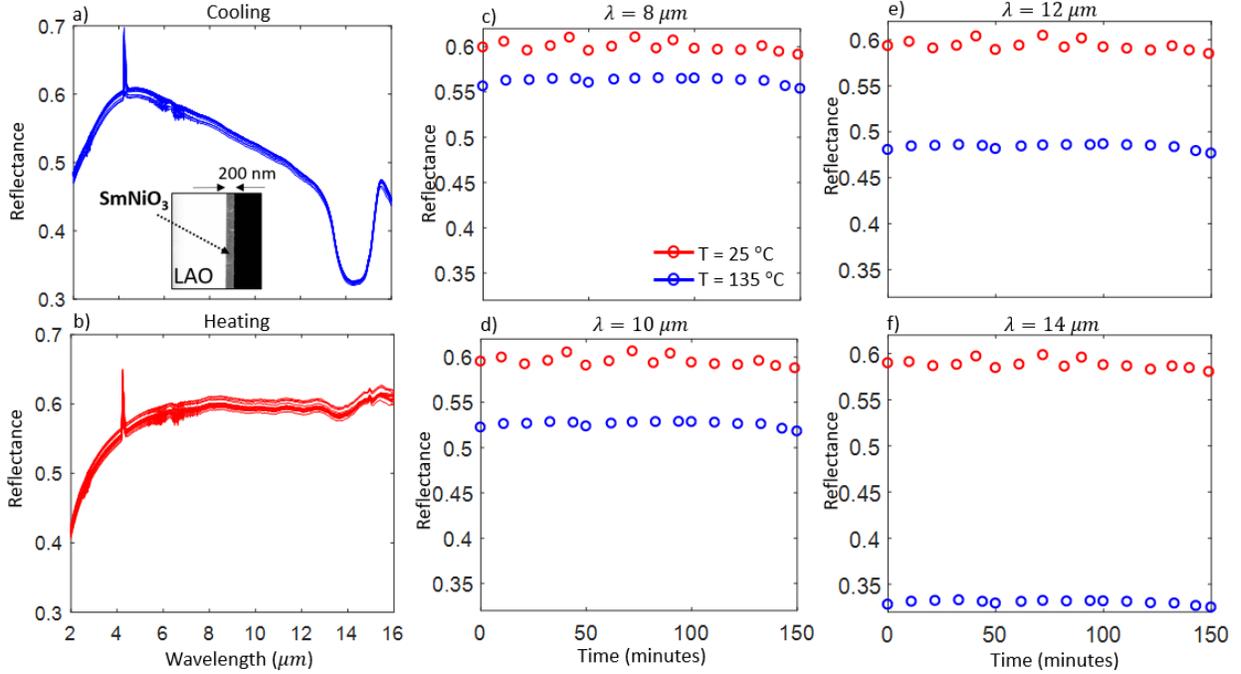

**Figure S2**. Reflectance of ~200-nm SmNiO$_3$ on LaAlO$_3$ across its phase-transition during both of **(a)** heating and **(b)** cooling. The inset shows a cross-sectional image of the sample. We thermally cycled the sample between 25 and 135 °C many times every two minutes for 150 minutes. The sharp peak is a result of atmospheric (CO$_2$) absorption. **(c-f)** show the reflectance at several different wavelengths ($\lambda$ = 8, 10, 12, and 14 μm).

## 2. Temperature-dependent X-ray diffraction

X-ray diffraction (XRD) measurements were carried out on an SmNiO$_3$ thin film deposited on a sapphire substrate, shown in Fig. 1. A wide-range XRD profile [Fig. S3(a)] shows clear diffraction peaks from orthorhombic SmNiO$_3$. Diffraction peaks from various planes are observed from SmNiO$_3$, indicating that this SmNiO$_3$ film possesses polycrystal structure, as expected. To investigate the structural evolution of SmNiO$_3$ across the IMT, the diffraction peak from the (110)/(002) planes of SmNiO$_3$ was measured *in-situ* upon heating [Fig. S3(b)]. No obvious shift of the diffraction peak from SmNiO$_3$ appears upon heating to 200 °C, which is consistent with the high-resolution neutron diffraction measurements reported previously [S3]. As expected, the IMT of SmNiO$_3$ is not driven by large lattice distortions but rather very subtle bond distortions [S3], which may contribute to its hysteresis-free and gradual phase change features observed in optical and electrical measurements. The lattice parameter change due to thermal expansion is within the resolution of the laboratory XRD and hence is a very subtle change.

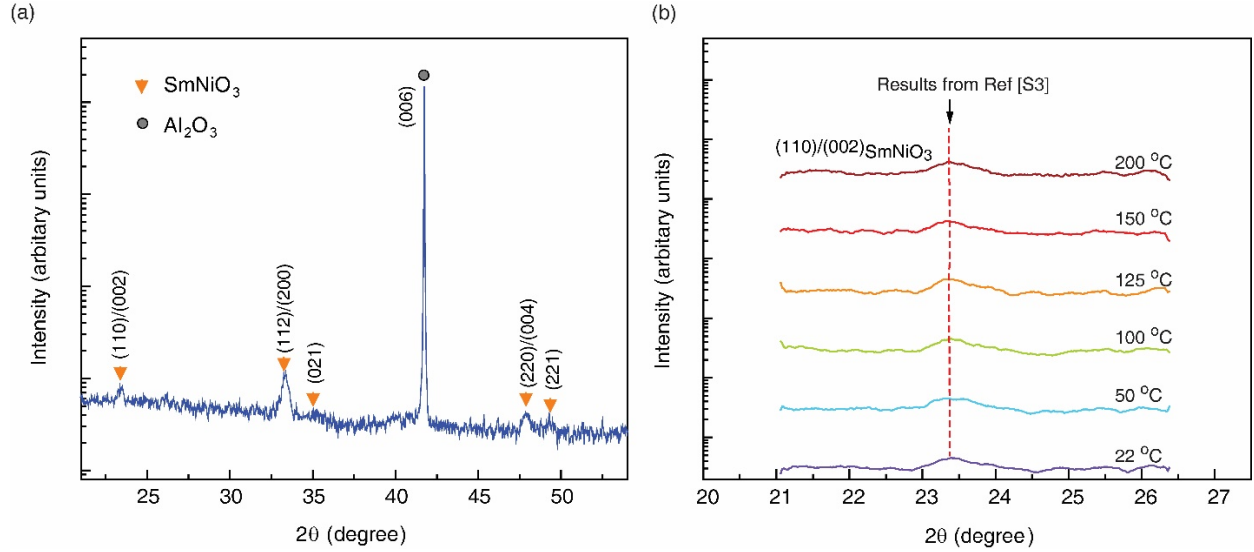

**Fig. S3. *In-situ* XRD measurement of SmNiO$_3$ thin film deposited on sapphire substrate.** (a) XRD profile of SmNiO$_3$ thin film measured at room temperature, where the diffraction peaks from SmNiO$_3$ and Al$_2$O$_3$ substrate are indicated. (b) In-situ XRD profile on (110)/(002) diffraction peak of SmNiO$_3$ upon heating. The red dashed line indicates the extracted 2θ value of SmNiO$_3$ from neutron powder diffraction reported in Ref [S3].

## 3. Comparison of ZDTE performance for SmNiO$_3$ on sapphire and LaAlO$_3$

Based on the characterized complex refractive index of SmNiO$_3$ and the optical properties of sapphire and LaAlO$_3$ from literature [S4], we estimated the ZDTE performance of SmNiO$_3$ films of different thickness on these two substrates. Our calculations show that both substrates work well, but superior performance is achieved with the sapphire substrate (Fig. S4).

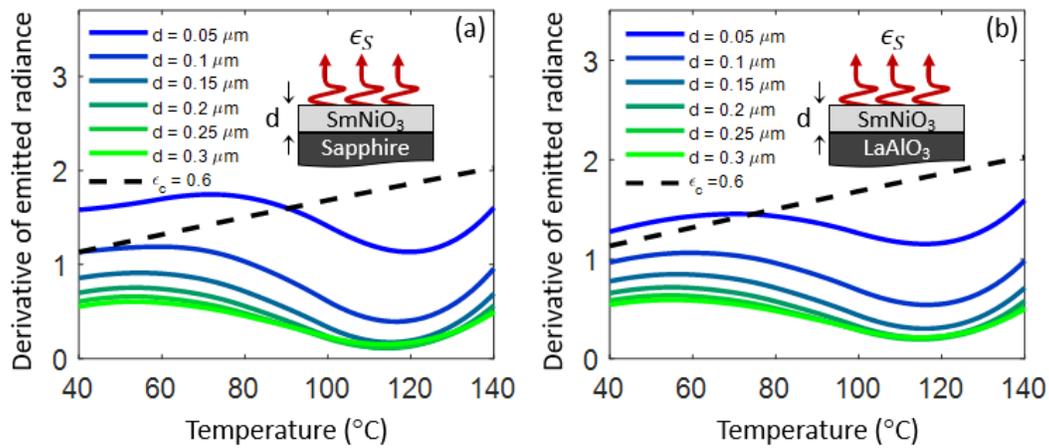

**Figure S4)** Calculated temperature derivative of the emitted radiance of an SmNiO$_3$ film with thickness $d$ from 50 nm to 300 nm on a (a) sapphire and (b) LaAlO$_3$ substrate. The dashed line yields the calculated radiance from a temperature-independent reference with a constant emissivity of 0.6.

## 4. X-ray absorption spectroscopy and x-ray photoemission electron microscopy

Nanoscale x-ray absorption spectroscopy (XAS) and x-ray photoemission electron microscopy (XPEEM) has been successfully used to image the metallic and insulating phase co-existence in the nickelate family [S5]. We performed XAS and XPEEM measurements on an epitaxial SmNiO$_3$ (001) thin film grown on a LaAlO$_3$ substrate, and observed no spatial features above the detector noise level. The spatially averaged XAS shows a disappearance of the double-peak structure near the Ni-L3 absorption edge across the IMT, corresponding to the melting of charge disproportionation (Fig. S5). The ratio of the absorption coefficients at 848 and 849 eV [$r = XAS\,(848\,eV)\,/\,XAS\,(849\,eV)$] is 1.02 in the insulating phase and 0.98 in the metallic phase. This value can therefore be used as a proxy for spatial identification of the local electronic state. The spatial mapping of the parameter $r$ can be obtained from the pixel-by-pixel ratio of two XPEEM images taken at 848 and 849 eV. Spatial maps of $r(x,y)$ over a field of view of 15 µm diameter are shown in Fig. S6, underscoring the absence of any domain texture or phase coexistence beyond the noise level, even for a temperature (120 °C) deep within the IMT.

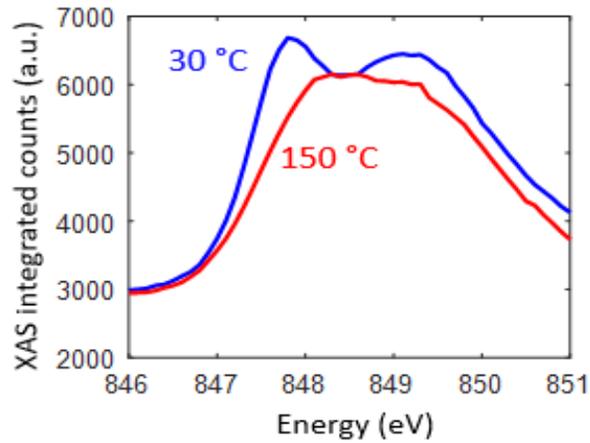

**Figure S5**. Spatially-averaged X-ray absorption spectra of SmNiO$_3$ (001) on an LaAlO$_3$ substrate at temperatures below and above the IMT. The XAS near the Ni L3 edge changes peak shape across the MIT transition. The ratio of the values at the two energies identified in the diagram is XAS(848)/XAS(849) = 1.02 at 30 °C and 0.98 at 150 °C.

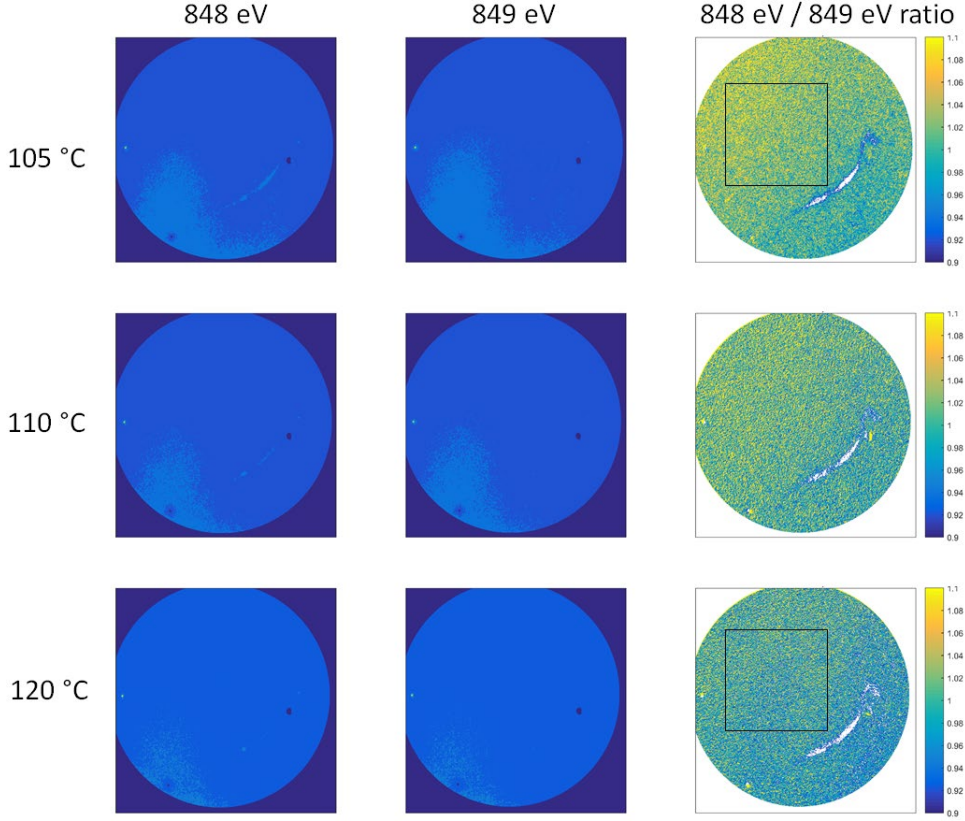

**Figure S6**. Left: XPEEM images of the same sample as in Fig. S4, at 848 eV and 849 eV. Images were taken with the same field of view, by positioning to surface defects and scratches. The color scale is normalized to the min and max of each map. Right: ratio between the 848 and 849 eV absorption maps. Data was acquired for three different temperatures throughout the IMT. The field of view (i.e., diameter of the circular aperture) is 15 µm. The maps have no recognizable spatial features other than a surface scratch on the detector (visible on the right), and detector noise.

## 5. Algorithm for extracting emissivity from direct-emission measurements

In general, one can describe the detected emission signal from an FTIR system as [S6]:

$$S_X(\lambda, T) = m(\lambda)[L_{BB}(\lambda, T)\varepsilon_X(\lambda, T) + B_X(\lambda, T)] \tag{S3-1}$$

where $S_X(\lambda, T)$ is the measured Fourier-transformed emission spectrum, $m(\lambda)$ is the system response, $\varepsilon_X(\lambda, T)$ is the temperature-dependent sample emissivity, $L_{BB}(\lambda, T)$ is the blackbody radiance, and $B_X(\lambda, T)$ is the background, which can also be sample dependent. To calibrate our FTIR, we performed temperature-dependent thermal-emission measurements using three samples: a vertically-aligned carbon nanotube forest (CNT) [S7] "blackbody" (with constant $\varepsilon \sim 0.95$ across the infrared spectral range), a fused-silica wafer, and a sapphire wafer.

Using Eq. (S3-1) and further analysis of our FTIR instrument [S8], we can obtain the emissivity of an unknown sample from a known reference $\alpha$ via:

$$\varepsilon_X(\lambda, T) = \varepsilon_\alpha(\lambda, T) \frac{S_X(\lambda, T)}{S_\alpha(\lambda, T)} \tag{S3-2}$$

Note that S3-2 does not work for all situations and systems; careful analysis is necessary to make sure that the emissivity is being extracted correctly. We measured the emissivity of our test samples (polished fused silica and sapphire wafers) by using this formula, and further confirmed that the extracted emissivity matches well with the value obtained by spectroscopic ellipsometry, Fresnel equations, and an application of Kirchhoff's law (Fig. 3(a) in ref. [S8]).

## 6. Angular response of our ZDTE

We numerically studied the angular dependence of the thermal radiance of our SmNiO$_3$/sapphire ZDTE. Based on our numerical analysis [Fig. S7(a-b)], the ZDTE effect persists for p-polarized emission up to an emission angle of about 40°, and for s-polarized emission for essentially every angle. Averaging the polarizations, the effect persists up to about 60° [Fig. S7(c)]. In our calculations, we incorporated the optical anisotropy in the sapphire substrate [S9] by using the transfer matrix method [S10], with the Fresnel coefficients generalized for birefringent media [S11]. The refractive indices of SmNiO$_3$ that we used were extracted by ellipsometry analysis, assuming that they were isotropic [Fig. 2(c, d)].

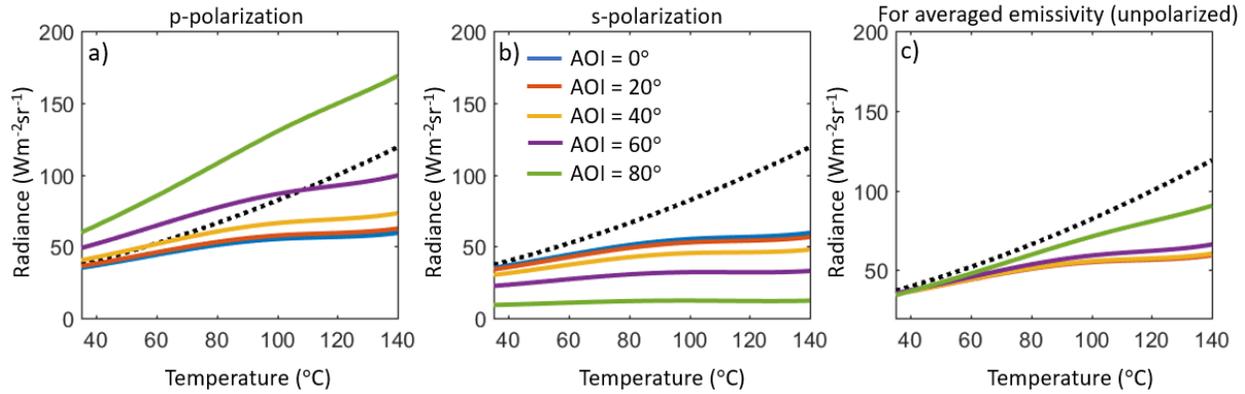

**Figure S7**. Calculated radiance versus temperature and angle for the structure characterized in Fig. 3(a) of the main text for **(a)** p-polarized, **(b)** s-polarized, and **(c)** polarization-averaged emissivity.

## 7. Temperature drop between the bottom and top surfaces of the substrate

Here, we provide our numerical analysis of the temperature drop due to the non-zero thermal conductivity and thickness of our emitter. We calculated the temperature distribution for a 0.5-mm-thick sapphire sitting on top of a 100 °C heater surface [Fig. S8(a)]. For this calculation, we solved one-dimensional heat equation ($\frac{\partial T(t,z)}{\partial t} = \frac{k}{\rho C_p}\frac{\partial^2 T(t,z)}{\partial z^2}$) along the sapphire wafer using the heat transfer parameters for sapphire from Table 1. From this calculation, we find out that the temperature drop is 0.02 °C.

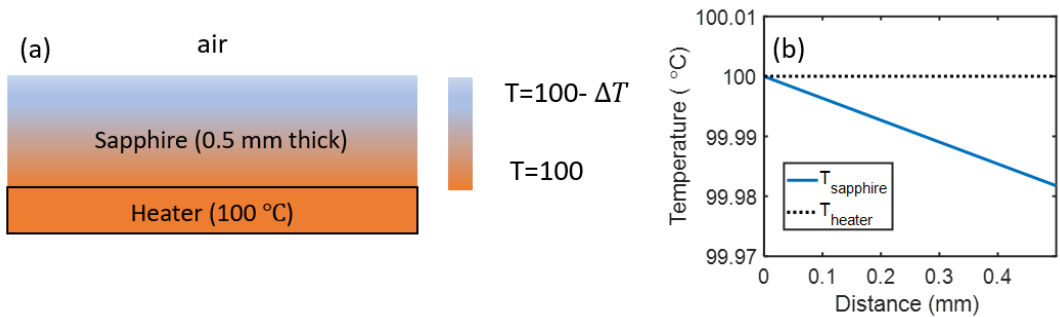

**Figure S8. (a)**: schematic of the heat-transfer simulation. **(b)** calculated steady state temperature profile for a 1-mm-thick sapphire on top of a 100 °C heater surface.

Table 1. Heat-transfer parameters for sapphire

| $k$ ( W/(m· °C) ) | $\rho$ (g/cm³) | $C_p$ ( J/(kg· °C) ) |
|---|---|---|
| 24* | 3.98 | 760 |

*The thermal conductivity is chosen to be the average value of perpendicular and parallel to the c-axis, though we note that the values are very close to each other.

## 8. Extending the temperature range of ZDTE

In this section, we show that it may be possible to achieve ZDTE over a different temperature range than what we demonstrated in the paper by leveraging other rare-earth nickelates with lower transition temperatures. As shown in Ref. [2], while the transition temperatures of SmNiO$_3$ and neodymium nickelate (NdNiO$_3$) are roughly 100 and −100 °C, respectively, Sm$_{0.75}$Nd$_{0.25}$NiO$_3$ has a transition temperature between these extremes [Fig. S9(a)]. While the optical properties of this or similar quaternary alloys have not yet been reported in the literature, the similarity in electrical behavior demonstrated in Fig. 1 of ref. [S2] suggests that the optical properties show some parallels to that of SmNiO$_3$. Therefore, we designed and simulated a zero-differential emitter for the temperature range of 45 to 80 °C using a 250-nm-thick Sm$_{0.75}$Nd$_{0.25}$NiO$_3$ thin film on a sapphire substrate, making the assumption that this alloy has the same spectral optical properties as SmNiO$_3$, but with a transition temperature 50 °C smaller than that of SmNiO$_3$. The calculated ZDTE effect is shifted to a temperature range centered at 60 °C [Fig. S9(b)].

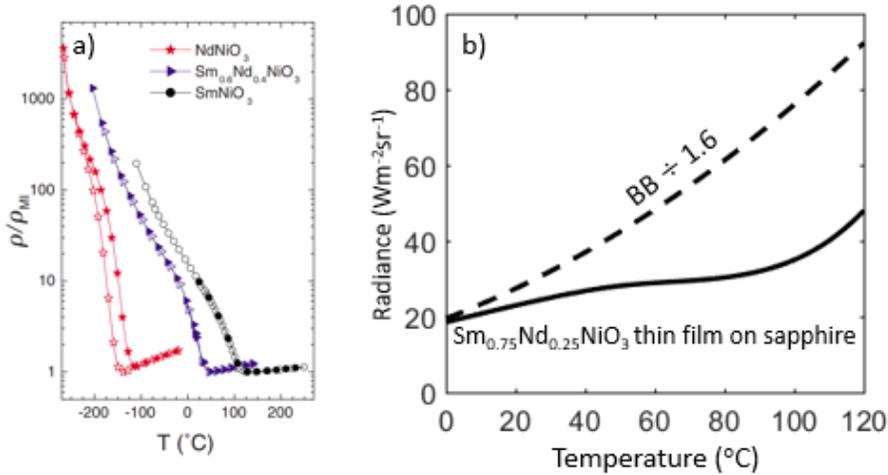

**Figure S9)** (a) Resistivity versus temperature for different rare-earth nickelates, including one quaternary alloy. This panel is reprinted from ref. [S2] with permission from APS. (b) Calculated emitted radiance for a 250-nm-thick Sm$_{0.75}$Nd$_{0.25}$NiO$_3$ thin film on sapphire, assuming that the Sm$_{0.75}$Nd$_{0.25}$NiO$_3$ film has the same optical properties as SmNiO$_3$ but with the transition temperature 50° lower than that of SmNiO$_3$.

**Supplementary references**

[S1]     Ha, S. D., Otaki, M., Jaramillo, R., Podpirka, A., & Ramanathan, S. (2012). Stable metal–insulator transition in epitaxial SmNiO3 thin films. *Journal of Solid State Chemistry*, *190*, 233-237.

[S2]     Girardot, C., Kreisel, J., Pignard, S., Caillault, N., & Weiss, F. (2008). Raman scattering investigation across the magnetic and metal-insulator transition in rare earth nickelate R NiO$_3$ (R= Sm, Nd) thin films. *Physical Review B*, *78*(10), 104101.

[S3]     Garcia-Munoz, J. L., Rodriguez-Carvajal, J., Lacorre, P., & Torrance, J. B. (1992). Neutron-diffraction study of R NiO 3 (R= La, Pr, Nd, Sm): Electronically induced structural changes across the metal-insulator transition. *Physical review B*, *46*(8), 4414.

[S4]     Lindermeir, E., Haschberger, P., Tank, V., & Dietl, H. (1992). Calibration of a Fourier transform spectrometer using three blackbody sources. *Applied optics*, *31*(22), 4527-4533.

[S5]     Mattoni, G., Zubko, P., Maccherozzi, F., van der Torren, A. J., Boltje, D. B., Hadjimichael, M., ... & Aarts, J. (2016). Striped nanoscale phase separation at the metal–insulator transition of heteroepitaxial nickelates. Nature communications, 7, 13141.

[S6]     Zhang ZM, Choi BI, Flik MI, Anderson AC (1994). Infrared refractive indices of LaAlO$_3$, LaGaO$_3$, and NdGaO$_3$. *Journal of the Optical Society of America B*.,11(11),2252-7.

[S7]     Mizuno, K., Ishii, J., Kishida, H., Hayamizu, Y., Yasuda, S., Futaba, D. N., ... & Hata, K. (2009). A black body absorber from vertically aligned single-walled carbon nanotubes. Proceedings of the National Academy of Sciences, 106(15), 0900155106.

[S8]     Xiao, Y., Shahsafi, A., Roney, P.J., Wan, C., Joe, G., Yu, Z., Salman, J. and Kats, M.A. (2019). Measuring Thermal Emission Near Room Temperature Using Fourier-Transform Infrared Spectroscopy. *Physical Review Applied,* 11(1),014026.

[S9]     Schubert, M., Tiwald, T. E., & Herzinger, C. M. (2000). Infrared dielectric anisotropy and phonon modes of sapphire. *Physical Review B*, *61*(12), 8187.

[S10]    S. J. Byrnes, Multilayer optical calculations, arXiv:1603.02720.

[S11]    Weber, M. F., Stover, C. A., Gilbert, L. R., Nevitt, T. J., & Ouderkirk, A. J. (2000). Giant birefringent optics in multilayer polymer mirrors. *Science*, *287*(5462), 2451-2456.